# Design and test of frequency tuner for CAEP high power THz free-electron laser [*]

MI Zhenghui (米正辉)[1;2;1] SUN Yi(孙毅)[2] PAN Weimin(潘卫民)[2] LIN Haiying(林海英)[2]
ZHAO Danyang (赵丹阳)[1;2] LU Xiangyang(鲁向阳)[3] QUAN Shengwen(全胜文)[3]
LUO Xing(罗星)[3] LI Ming(黎明)[4,5] YANG Xingfan(杨兴繁)[4,5] WANG Guangwei(王光伟)[2]
DAI Jianping(戴建枰)[2] LI Zhongquan(李中泉)[2] MA Qiang(马强)[2] SHA Peng(沙鹏)[2]

(1 University of Chinese Academy of Sciences, Beijing 100049, China; 2 Institute of High Energy Physics, CAS, Beijing 100049, China; 3 State Key Laboratory of Nuclear Physics and Technology, Peking University, Beijing 100871, China; 4 Institute of Applied Electronics, CAEP, P.O. Box 919-1015, Mianyang 621900, China; 5 Terahertz Research Center, CAEP, Mianyang 621900, China)

**Abstract**：Peking University is developing a 1.3 *GHz* superconducting accelerating section for China Academy of Engineering Physics (CAEP) high power THz free-electron laser. A compact fast/slow tuner has developed by Institute of High Energy Physics (IHEP) for the accelerating section, to control Lorentz detuning, beam loading effect, compensate for microphonics and liquid Helium pressure fluctuations. The tuner design, warm test and cold test of the first prototype are presented.

## 1. Introduction

CAEP is building a 8 *MeV* high power THz free-electron laser, which adopt two 4-cell superconducting accelerating cavities to accelerate electron. The cavity operating at a frequency of 1.3 *GHz* and a temperature of 2 *K*. It works at continue waves (CW) mode, while the maximum current is 5 mA. As figue 1 is the schema of superconducting accelerating section[1].

As with other pulsed superconducting cavities, for example the 1.3*GHz* 9-cell cavity for ILC, each 4-cell cavity will be equipped with a slow tuner to compensate for static detuning and a fast tuner to compensate for the dynamic detuning due to the Lorentz force, beam loading effect, microphonics and liquid Helium pressure fluctuations. The slow tuner for the 4-cell cavity uses a normal stepper motor while the fast tuner employs piezo actuators[2][3][4].

The 4-cell cavity are considerably stiffer than a 9-cell cavity for ILC and FEL, so the tuner must be capable of applying stronger force to the cavity flanges than the 9-cell cavity tuner. At the same time, the tuner will be mounted to internal magnetic shield so need to strictly limit the tuner remanence. And the 2 *K* heat leakage of the tuner is strictly limited also. Furthermore the planned potics for the superconducting accelerating section provides only limited space for a tuner. Therefore, because of these reasons the design of a suitable tuner is very challenging.

[*] Supported by the "500MHz superconducting cavity electromechanical tuning system", under Grant No. Y190KFEOHD
1) E-mail: mizh@ihep.ac.cn

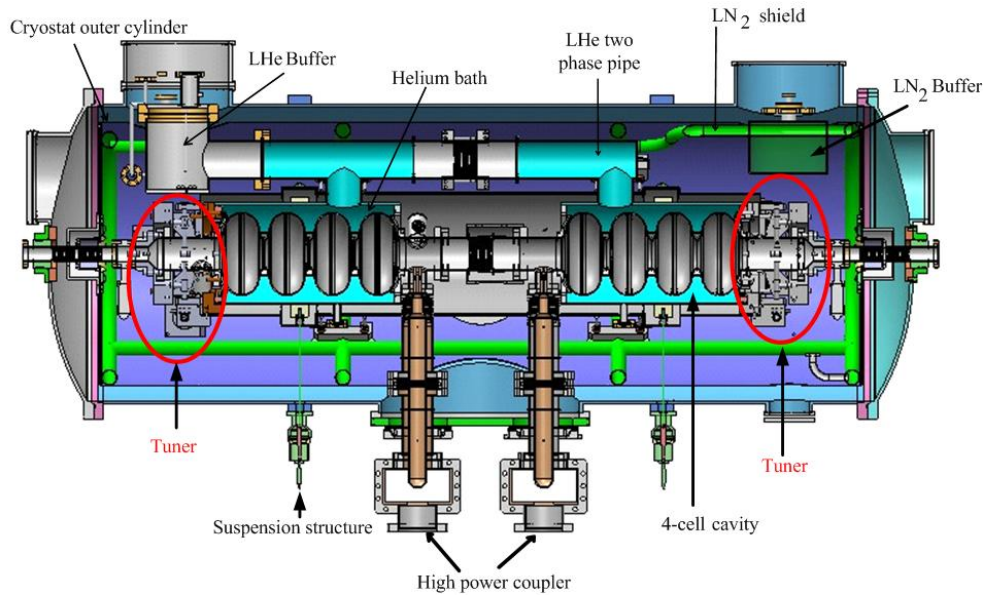

Figure 1 Whole schema of superconducting accelerating section

**2. Tuner requirements**

The tuner must be able to compensate for the following effects[4].

- Static variations in the resonant frequency of the cavities due to manufacturing tolerances.

- Dynamic detuning of the cavities by the Lorentz force during the RF pulse (the accelerating section works at pulse mode in the start stage).

- Slow variations of the cavity resonant frequency due to the helium pressure fluctuations, beam loading and microphonics.

The primary factors that drive the design of the tuner as follows:

◆ The spring constant of the 4-cell cavity (with helium bath) is about 16 *N/um*.

◆ The tuning sensitivity of the cavity is 770 *Hz/um*.

◆ The tuner install space is be limited by pickup and HOM port.

◆ The cavity requires the magnetic field at any position of the central line of the tuner less than 10 million Gauss (As figure 9 the tuner is wrapped up by magnetic shielding).

◆ The heat leakage of the two tuners should less than 0.3 *W* due to the limits of the 2 *K* cryogenic system.

Given the manufacturing tolerances for the cavities, the tuner must be able to statically retune the cavity over a range of 800 *kHz* centered on the nominal operating frequency of 1.3 *GHz*.

Based on simulations, at an accelerating gradient 13 *MV/m*, the Lorentz force is expected to detune the cavity by about 200 *Hz*. This detuning is significant when compared to the loaded width of the cavity resonance of 260 *Hz* (@$Q_L$=5×$10^6$). If the Lorentz force detuning is not compensated, additional RF power will be required to achieve the desired accelerating gradient.

About 50 *Hz* detuning will be lead by 5 *mA* beam loading, when the cavity working at CW mode with 4 *MV* cavity voltage.

At the planned operating temperature of 2 *K* slow fluctuations in the pressure of the surrounding liquid He bath can change the resonant frequency of the cavity over periods of several minutes. An ANSYS model estimated the sensitivity to be: $\Delta f / \Delta P$ =100 *Hz/mbar*.

The cryogenic system will be able to regulate the helium pressure to within ±0.1 *mbar*. Pressure variations of this magnitude can shift the resonant frequency by up to ±10 *Hz*.

Furthermore, in the event that the cavity should fail, the tuner must be able to statically detune the cavity by about 200 *kHz* to limit any beam-cavity coupling (About 55 *kW* power output from 5 *mA* beam at resonant frequency).

## 3. Tuner design

The major elements of the tuner are shown in Figure 2. The tuner is amounted on the pickup end. Four annular flanges are fixed with the cavity brim. Two bow beams connect the front flange and annular flange together. A stepper motor controls the position of each roller way via a harmonic drive assembly and screw gear mechanism. Gear set was chosen to provide a reduction ratio of 3:1. The wedge provide a ratio of 15:1 between the displacements of the roller way and the front flange of motor tuner[6].

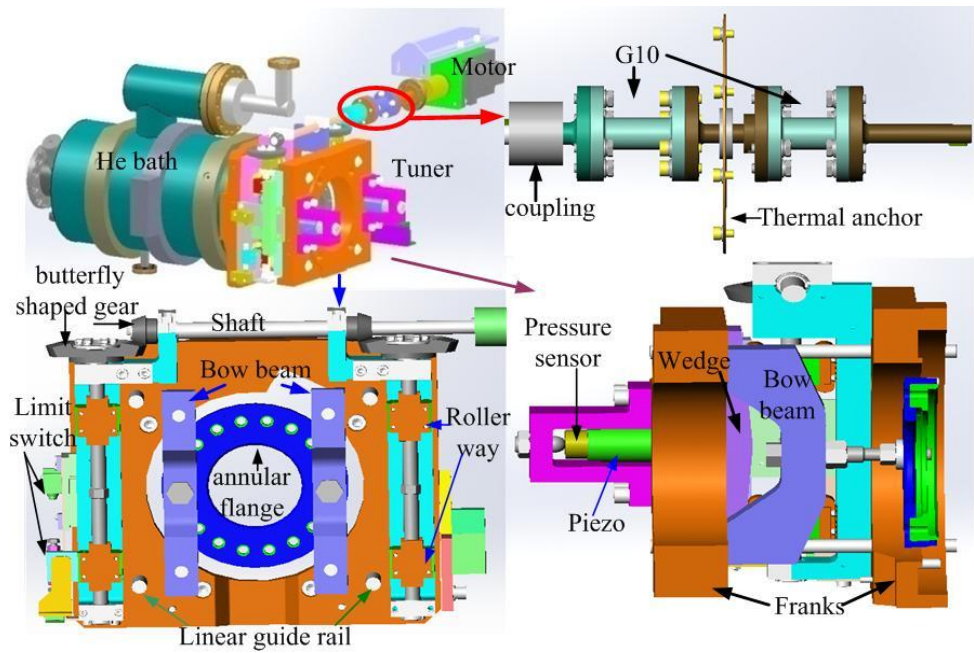

Figure 2 Details of 4cell cavity tuner design

Motion of the wedge is transmitted to the front flange of the motor tuner by two piezo actuators. Through two bow beams fixed on the front flange the cavity is stretched. To accommodate the large forces required to tensile the cavity two actuators were used at each end.

To deliver the required static tuning range of 260 *kHz* to 800 *kHz*, the motor tuner must displace the cavity by about 351 to 1039 *um*. The static load on each piezo will change by ~5.5 *kN* or 15.7% of piezo blocking force. To ensure that the load on the piezo remains between 10% and 50% of the blocking force, the tuner is designed to always apply tensile force to the cavity flanges across the entire tuning range.

Piezo actuators were selected to provide the cold stroke of 4 *um*. To avoid shear forces on the actuators and to simplify the connection with the arms, the piezo actuators that have been encapsulated in custom stainless steel housing was chosen.

On the roller way two limit switches are installed, which are designed to limit the stroke of motor tuner.

During warm test or cold operation, the force applied to the cavity flanges can be monitored using two pressure sensors installed at the end of piezo acturators.

The tuner has been designed so that it can be assembled and tested as an independent unit prior to being mounted on the cavity. Once testing is completed, the assembly is installed on the helium vessel and initial piezo preload are set using adjustment screws.

As below figure 3 (a) shown the mechanical deformation of the tuner is about 53 *um* under the load of 21300 *N*. The ratio of deformation of tuner and cavity is 1: 25. The max stress of the tuner is about 79 *MPa* shown as figure 3 (b), which is meets the mechanical safety requires.

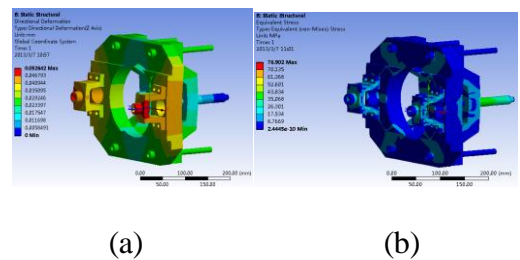

(a)          (b)

Figure 3 Simulation of stress and deformation

## 4. Test results

Figure 4 shows the test stand of tuner. A prototype tuner was installed on the helium vessel of the 4-cell cavity. The performance of tuner was test under room temperature and 80 *K* liquid nitrogen temperature[5].

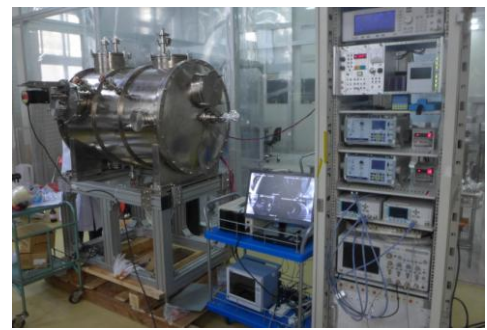

Figure 4 Test stand of tuner

During the stage of cooling-down, the frequency and load of the cavity were changing as figure 5 shown. The preload is 130 *kg*, while after the cavity was evacuated the preload decreased to 60 *kg*. During the decreasing of the temperature the load of the cavity would increase to 254 *kg*. The frequency of the cavity change about 1.78 *MHz* during cooling-down, and the final frequency is over working frequency (1.3 *GHz*) at 80 *K* temperature. So it should be tuned to a certain range that between 1.29705 to 1.29739 *GHz* at pretuning stage, in order to keep the frequency within a suitable tuning range and the load force of the cavity is between 17% to 50%.

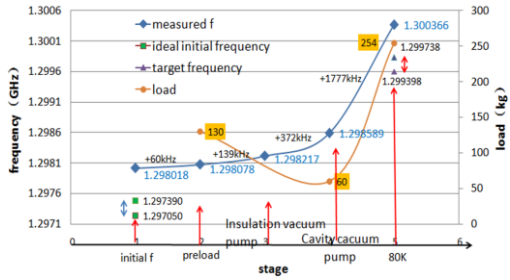

Figure 5 Load and frequency during cooling-down

### 4.1 Motor tuner test result

To evaluate the performance of the tuner at room temperature, the resonant frequency of the cavity was monitored using a networking analyzer as the stepper motor was operated. During motor operation, the forces on the cavity flanges were monitored by the pressure sensors and the displacement of the cavity was measured with dial gauges.

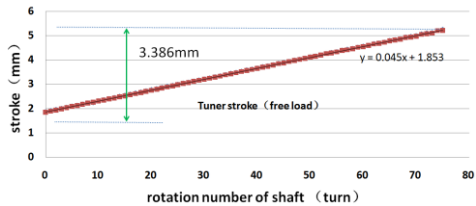

(a)

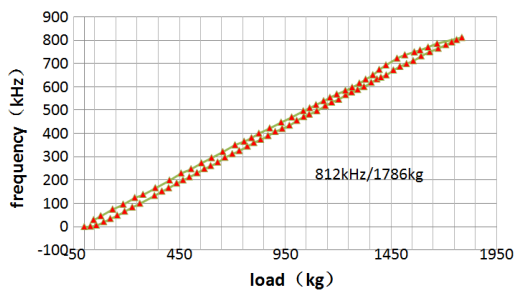

(b)

Figure 6 Tuning range of motor tuner

Figure 6 (a) is room temperature test result of motor tuner. The max stroke of the tuner is about 3.386mm. At 80 *K* liquid nitrogen temperature, the tuning max range of motor tuner can reach 812 *kHz* as figure 6 (b) shown. The max force is about 25.7% of the blocking force.

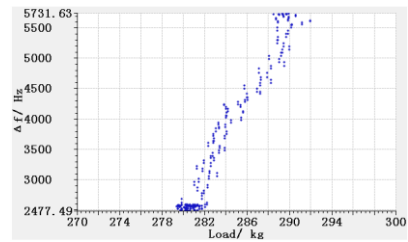

Figure 7 Resolution of motor tuner

Use slow keys of tuner controller to control the motor, the minimum amount of frequency changing of the cavity that measured is about 50 ~ 60 *Hz* as shown in figure 7. Due to the IQ phase has 0.06 degrees of jitter, which corresponding to the frequency jitter is 52 *Hz*, so the resolution of the motor tuner measured was limited by the precision of phase detection.

### 4.2 Piezo tuner test result

The performance of the piezo tuner was evaluated by measuring the stroke and tuning range.

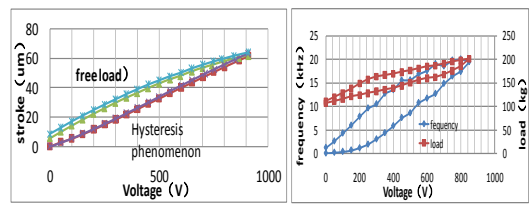

(a)          (b)

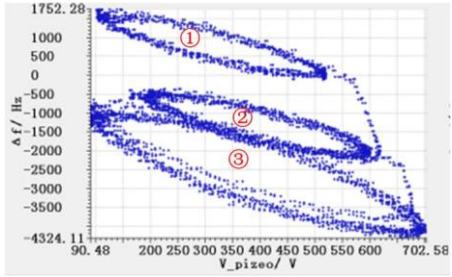

(c)

Figure 8 Test curve of piezo tuner

As figure 8 (a), the stroke of the piezo tuner is about 60 *um* at free load and room temperature, corresponding frequency change of the cavity should be 42 *kHz*, while in fact the frequency change of the cavity only about 20 *kHz* as shown in figure 8 (b) shown. Nearly half of the displacement was consumed by mechanical clearance. Therefore, the assembly process of the tuner required each part is fixed tightly.

The piezo tuner was tested at 80 *K* liquid nitrogen temperature as figure 8 (c) shown. Sine signal was magnified by high voltage amplifier to drive the piezo tuner. As curve ① the tuning rang is about 1.8 *kHz* at the offset voltage of 300 *V* and scan voltage range ±200 *V*, curve ② shows the tuning rang is also 1.8kHz at the offset voltage of 400 *V* and scan voltage range ±200 *V*, while curve ③ displays the tuning rang is about 3.3 *kHz* at the offset voltage of 400V and scan voltage range ±300 *V*. So the tuning range of piezo tuner meets the requires at offset voltage of 400 *V* and scan voltage range ±300 *V*.

**4.3 Axial magnetic field measurement**

The tuner and cavity were wrapped by magnetic shielding as figure 9 shown. Because the value of residual magnetism has great effect on superconducting cavity $Q_0$, so needs to limit the axial magnetic field of the tuner.

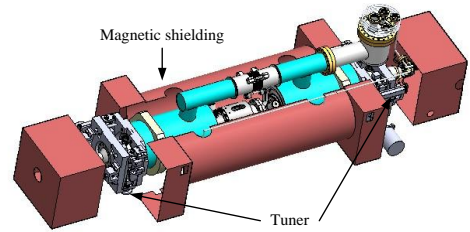

Figure 9 Magnetic shielding of the cavity and tuner

The mechanical structure of motor tuner is made of 304 stainless steel. The magnetic field of the tuner was measured by a Gauss magnetic measurement instrument in a vertical test Dewar, as shown in figure 10.

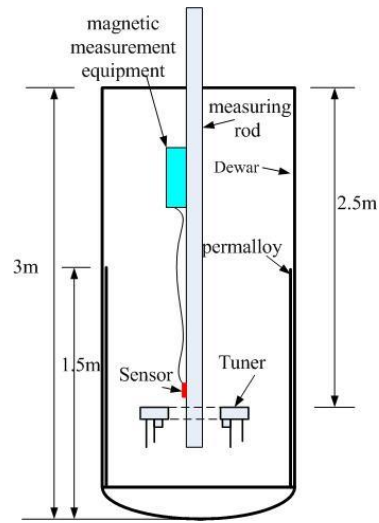

Figure 10 Schematic diagram of magnetic field measurement of tuner

The test results of magnetic field as figure 11 shown. The upper end of the Dewar as zero position.

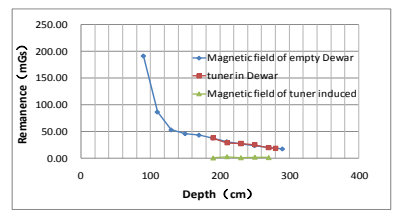

Figure 11 magnetic field measure results

The inner intensity of magnetic field

of Dewar decreases with the increasing of depth. From the depth of 0 cm to 130 *cm* the intensity of magnetic field decreased rapidly, while the depth over 130 *cm* it decreased slowly and tended to be stable. The strength of the magnetic field of the tuner leads less than 3 *mGs*, which meets the requirements of axial magnetic field of the tuner no more than 10 *mGs*.

Important tuner parameters are summarized in table 1 as below:

Table 1: Tuner Design and Performance Parameters

| Parameter | Designed | Measured |
|---|---|---|
| Cavity Spring Constant(*N/mm*) | 16 | ~15 |
| Cavity Sensitivity(*Hz/N*) | 770 | ~700 |
| Piezo Sensitivity(*Hz/V*)(room temperature) | 46.2 | 23.5 |
| Gear set | 1:3 | 1:3 |
| Wedge | 1:15 | 1:15 |
| Maximum Piezo Load(*N*) | 16000 | 17860 |
| Motor tuning range(*kHz*) | 800 | 812 |
| Piezo tuning range (*kHz*) | ≥2 | 3.3 |
| Axial magnetic field of tuner | ＜10 *mGs* | ~3 *mGs* |
| Heat leakage | ＜0.3W | -- |
| Piezo Preload(*N*)(initial piezo load) | 130 | 130 |
| Number of piezo | 2 | 2 |
| Harmonic Drive Ratio | 1:50 | 1:50 |

## 5. Conclusion

The tuning range and axial magnetic field of tuner have satisfied the design requirements through room temperature and 80*K* liquid nitrogen temperature test. Due to the limit of test condition, the dynamic performance and resolution of tuner didn't do. The work performance of tuner needs further research.

## 6. Acknowledgment

The authors wish to thank the members of IHEP RF and Cryogenic group for their perfect cooperation during our work. Special thanks go to PKU and CAEP experts for their help to us.